\documentclass{article}
\newcommand{\be}{\begin{equation}}
\newcommand{\ee}{\end{equation}}
\newcommand{\bea}{\begin{eqnarray}}
\newcommand{\eea}{\end{eqnarray}}
\newcommand{\ba}{\begin{array}} \newcommand{\ea}{\end{array}}

\usepackage{amscd,amsmath,amssymb}

\begin{document}
\thispagestyle{empty}
\begin{center}
{\Large\bf Isospin particle  systems on quaternionic projective
spaces}\\[7mm]

{\large Stefano Bellucci$^a$, Sergey Krivonos $^b$, Armen
Nersessian$^{c}$, Vahagn Yeghikyan$^{a,c}$ }
\\[3mm]

$\;^a${\sl INFN-Laboratori Nazionali di Frascati,Via E. Fermi 40,
00044,
Frascati, Italy}\\
 $\;^b$ {\sl
Bogoliubov Laboratory of Theoretical Physics, JINR,
141980 Dubna, Russia}\\
 $\;^c${\sl Yerevan State University, 1 Alex
Manoogian St., Yerevan, 0025, Armenia}
\end{center}

\begin{abstract}\noindent
We construct the isospin particle system on $n$-dimensional
quaternionic projective spaces in the presence of BPST-instanton
by the reduction from the free particle on $(2n+1)$-dimensional
complex projective space.  Then we add to this system  a
"quaternionic oscillator potential" and show, that this
oscillator-like system is superintegrable. We show, that besides
the analogs of quadratic constants of motion of the spherical
(Higgs) and $\mathbb{C}P^n $- oscillators,  it possesses the
third-order constants of motion, which are functionally
independent from the quadratic ones.
\end{abstract}
\setcounter{page}{1}
\setcounter{equation}{0}
\section{Introduction}
Hopf maps play a distinguished role in theoretical physics,
appearing, sometimes in a hidden way, in most of the key models.
However, even the constructions, related with the second Hopf map,
particularly, quaternionic projective spaces $\mathbb{H}P^n$, are
not properly studied or/and used. In fact, explicitly quaternionic
projective spaces appear in the construction of the
multi-instanton of self-dual Yang-Mills theory only \cite{atiah}.
Even the classical and quantum mechanical systems on quaternionic
projective spaces (except systems on $\mathbb{H}P^1$ i.e. the
four-dimensional sphere) were not paid enough attention. On the
other hand, there is no doubt, that on these spaces one can easily
construct the integrable systems of isospin particles interacting
with instantons: due to the existence of the well-known fibration
$S^2\to\mathbb{C}P^{2n+1}\to\mathbb{H}P^n$, the inclusion of
instanton fields should not destroy the symmetries of the
$Sp(n)$-invariant systems on $\mathbb{H}P^n$. This is similar to
the well-known preservation of the symmetries of $U(N)$ invariant
systems on complex projective spaces after inclusion of the
constant magnetic field, which reflects the existence of the
fibration $S^1\to S^{2n+1}\to\mathbb{C}P^n$ related with the first
Hopf map. Moreover, it is obvious, that on $\mathbb{H}P^n$ one can
define the $sp(n+1)$-invariant "quaternionic Landau problem", i.e.
a free particle interacting with a constant (BPST) instanton
field, which is the quaternionic analog of the "Landau problem" on
$\mathbb{C}P^n$: a $su(n+1)$-invariant system of particles
interacting with a constant magnetic field. The simplest,
one-dimensional quaternionic Landau problem on $\mathbb{H}P^1=S^4$
\cite{horvathy} has been used previously for developing the model
of the "four-dimensional Hall effect" \cite{science}, and, by this
reason it attracted much attention (see, e.g.  \cite{4Hallothers}
and the brief review \cite{Karabali}). Nevertheless, all these
studies were restricted to the systems on $\mathbb{H}P^1=S^4$, and
there was no attempts to consider even the higher-dimensional
quaternionic Landau problem. Though, technically this should not
be a difficult problem, since the fibration
$S^2\to\mathbb{C}P^{2n+1}\to\mathbb{H}P^n$ allows one to construct
the lift (or reformulate it) from the  free particle systems on
the complex projective space $\mathbb{C}P^{2n+1}$. For the  $n=1$
case this fibration was widely explored in the study of the
four-dimensional Hall effect, while the  $S^4$-Landau problem in
itself was explicitly constructed by the Hamiltonian reduction of
the free particle on $\mathbb{C}P^{3}$ in \cite{casteill}. Below
we will  fill the mentioned gap, presenting the detail description
of the Hamiltonian reduction of the free particle on
$\mathbb{C}P^{2n+1}$ to the quaternionic Landau problem on
$\mathbb{H}P^{n}$ ({\sl see the Third Section}).  Besides, we will
present the superintegrable analog of the oscillator  on
quaternionic projective spaces, which respects the inclusion of a
constant instanton field. In contrast with spherical (Higgs)
\cite{higgs} and $\mathbb{C}P^n$ \cite{cpnosc} oscillators, whose
hidden symmetries are of the second order in momenta, our model
has additional constants of motion, which are of the third order
in momenta({\sl see the Fourth Section}). In the {\sl Second
Section} we describe the fibration
$S^2\to\mathbb{C}P^{2n+1}\to\mathbb{H}P^n$.

\section{$\mathbb{C}P^{2n+1}\to \mathbb{H}P^n$ fibration}

In this Section we formulate the fibration and define the
mathematical objects we are going to deal with.

First, let us notice, that the definition of projective spaces
define an infinite series of fibrations, the natural projections
of which are called Hopf maps. Indeed, by definition the
projective space over a field  $\mathbb{F}$
($\mathbb{F}=\mathbb{C,H}$) is the set of all the lines through
the origin. A natural chart on this manifold is given by the
formula: \be q_i^{(k)}=v_iv_k^{-1},\quad i,k=1,\ldots,n+1, \ee
where $v$ define coordinates of the corresponding
$\mathbb{F}^{n+1}$ and $q_i^{(k)}$ is the $i$-th coordinate of the
$k$-th chart of the $\mathbb{F}P^n$. This maps define two infinite
families of tautological fibrations \be S^{2^N-1}\to S^{2^N
(n+1)-1}\to \mathbb{F}P^n,\label{taut}\ee where $2^N$($N=1,2$) is
the dimensionality of the corresponding field $\mathbb{F}$. Each
first element of these families is the famous Hopf fibration of
sphere over sphere: \be S^1\to S^3 \to S^2, \quad S^3\to S^7\to
S^4.\label{hpf} \ee

In our research we are interested in the projectivization of the
second fibrations in \eqref{taut}. Namely, it is possible to
project the total space and the fiber of \eqref{taut} with $N=2$
using the projection of the corresponding fibration with $N=1$ as
a map, so that it will not affect the base. After projectivization
we will arrive to the following fiber bundle: \be
S^2\to\mathbb{C}P^{2n+1}\to\mathbb{H}P^n\label{mnprj} \ee

Now, let us pass to the explicit construction of these fibrations.
We start from the $2n+2$-dimensional complex plane
$\mathbb{C}^{2n+2}\simeq \mathbb{H}^{n+1}$ with complex
coordinates $\lambda$ or quaternionic ones:
$v_i=\lambda_{2i-1}+j\lambda_{2i}$( $i=1,\ldots,n+1$ ). By
definition the coordinates \be q_\alpha=v_\alpha
v_{n+1}^{-1}\equiv v_\alpha\frac{\bar v_n}{||v_n||^2},\quad
\alpha=1,\ldots,n. \ee define a chart on the quaternionic
projective space $\mathbb{H}P^n$.

The inverse formulas look as follows:
\be
v_\alpha=q_\alpha v_{n+1}=q_\alpha(\lambda_{2n+1}+j\lambda_{2n+2})=\lambda_{2\alpha-1}+j\lambda_{2\alpha}
\ee

Multiplying the last equation by $\lambda_{2n+2}^{-1}$ one finds

\be q_\alpha(z_{2n+1}+j)=z_{2\alpha-1}+jz_{2\alpha}.\label{prj}
\ee where the quantities $z_r=\lambda_r/\lambda_{2n+2}$
$(r=1,\ldots,2n+1)$ define a chart on the complex projective space
$\mathbb{C}P^{2n+1}$. It is clear, that any  coordinate of
$\mathbb{C}P^{2n+1}$ by itself defines a chart on a
$\mathbb{C}P^1\simeq S^2$. In particular, one can consider as such
the last coordinate $z_{2n+1}$.

We can rewrite \eqref{prj} in the following form:
\be
z_{2\alpha-1}+jz_{2\alpha}=q_\alpha(u+j),\quad z_{2n+1}=u.\label{trans}
\ee
In this form those relations define a natural projection of the fibration \eqref{mnprj}

The form of transition functions can be easily found from the construction described above.

For our further consideration it is convenient, instead of the
quaternionic coordinates $q$, to use complex coordinates $w$ which
we introduce by the following formula:

\be q_\alpha=w_{2\alpha-1}+jw_{2\alpha}\ee
In these coordinates \eqref{trans} takes the following form:
\be
z_{2\alpha-1}=w_{2\alpha-1}u-\bar w_{2\alpha},\quad z_{2_\alpha}=w_{2\alpha}u+\bar w_{2\alpha-1},\quad  z_{2n+1}=u.\label{ctr}
\ee

 In order to unify the first two expressions we introduce a matrix $\Omega$ by the following formula:
\be
(\Omega_{\mu\nu})=\left(\begin{array}{cccccc}
              \varepsilon & 0 & 0 & 0 &\ldots \\
          0 &  \varepsilon & 0 & 0 &\ldots \\
          0 & 0 & \varepsilon & 0 &\ldots \\
          \ldots & \ldots & \ldots & \ldots &\ldots \\\\
             \end{array}
\right),\quad \varepsilon=\left(\begin{array}{cc}
              0 & 1\\
          -1 & 0\\      \end{array}
\right).
\ee

With this matrix we can rewrite \eqref{ctr} in the following form:
\be
z^\mu=uw^\mu+\Omega^{\mu\nu}\bar w_\nu,\quad z_{2n+1}=z^{2n+1}=u\quad \mu,\nu=1,\ldots,2n.\label{ctro}
\ee

{\bf Remark.} From this point we will make a difference between
the upper and lower indices. We define $w^\mu$ with upper index,
while its complex conjugate has a lower one: $\bar w_\mu$. The
rule of raising and lowering the indices is given via the matrix
$\Omega_{\mu\nu}$ and its inverse $\Omega^{\mu\nu}$: \be
\Omega_{\mu\nu}\Omega^{\nu\lambda}=\delta^\lambda_{\mu} \ee Thus,
we define \be w_\mu=\Omega_{\mu\nu}w^\nu,\quad \bar
w^\mu=\Omega^{\mu\nu}\bar w_{\nu} \ee The contraction is done, as
usual, between upper and lower indices. So, \be z\bar z=z^\mu\bar
z_\mu=-z_\mu\bar z^\mu \ee

Now, using the above established relations between
inhomogeneous coordinates of complex and quaternionic spaces, let us
relate metrics on these spaces.

It is known that the natural metric on $S^{2^N (n+1)-1}$ induces the
Fubini-Study metric on the corresponding projective space:
\be
ds^2=\frac{dz\bar dz}{1+z\bar z}-\frac{(\bar z dz)(zd\bar
z)}{(1+z\bar z)^2}.\label{mcpn}
\ee
 The non-degenerate
transformation \eqref{ctro} defines a connection on the fibration
\eqref{mnprj}. Indeed, replacing the coordinates $z_i$ with $(q,u)$
transforms the Fubini-Study metric on $\mathbb{C}P^{2n+1}$ as
follows:
\be ds^2=\frac{dq\bar dq}{1+q\bar q}-\frac{(\bar q
dq)(d\bar q q)}{(1+q\bar q)^2}+\frac{(du+A)(d\bar u+\bar
A)}{(1+u\bar u)^2}, \label{hpn}\ee
where \be A=j\left.\frac{(\bar
u-j)\bar q dq(u+j)}{1+w\bar w}\right|_\mathbb{C}\equiv\frac{u(\bar
wdw-wd\bar w)-(\bar w\Omega d\bar w)-u^2(w\Omega d w)}{1+w\bar
w}.\label{vp} \ee
Here $q|_\mathbb{C}\equiv 1/2(q-\imath q\imath)$
denotes the complex part of the quaternion  $q$.
%
In complex coordinates $w$ the metrics of $\mathbb{H}P^n$ reads
\be
g_\mu{}^\nu=\frac{\delta_\mu^\nu}{1+w\bar w}-\frac{\bar w_\mu w^\nu +w_\mu \bar w^\nu}{(1+w\bar w)^2}.\label{hpncp}
\ee

The complex projective space is a Riemannian symmetric space.
Indeed, each sphere of the total space in \eqref{mnprj} can be
represented as a coset space \be U(n+1)/U(n)\simeq S^{2n+1}, \ee
Reducing this by the global factor $U(1)$, we find: \be
SU(n+1)/U(n)\simeq \mathbb{C}P^{2n+1}. \ee Thus, the isometries of
the complex projective space form the $su(n+1)$ algebra. These
isometries are defined, in the given parametrization, by the
following vector fields \be\label{subos1} R_i= \partial_i +\bar
z_i (\bar z\bar \partial),\quad J_i{}^j = \imath\left( z^j
\partial_i -\bar z_i \bar \partial^j\right)+
\imath\delta_i^j\left( (z \partial) -(\bar z\bar \partial)\right).
\ee These are all we need to know about the transformation of
coordinates of the $\mathbb{C}P^{n}$.

Now, we are ready to construct a mechanical system of a particle on $\mathbb{H}P^n$
in the vector-potential \eqref{vp}  by the reduction of the free particle on $\mathbb{C}P^{2n+1}$.

\setcounter{equation}{0}
\section{ Landau problem on $\mathbb{H}P^n$ from a free particle on $\mathbb{C}P^{2n+1}$ }
Let us show, that the free particle on $\mathbb{C}P^{2n+1}$ is
immediately reduced to the particle on $\mathbb{H}P^n$ in the
presence of BPST instanton field (which is natural to call the
``Landau problem on $\mathbb{H}P^n$").

In accordance with (\ref{mcpn}), the free particle on
$\mathbb{C}P^{2n+1}$ is defined by the Lagrangian \be
L_0=\frac{{\dot{\bar z}}\cdot{\dot z}}{1+z\bar
z}-\frac{({\dot{\bar z}} z)({\dot z}\bar z)}{(1+z\bar z)^2}. \ee
In terms of \eqref{ctro} it reads \be L=\frac{\dot{q}\dot{\bar
q}}{1+q\bar q}-\frac{(\bar q \dot{q})(\dot{\bar q} q)}{(1+q\bar
q)^2} +\frac{(\dot{u}+A)(\dot{\bar u}+\bar A)}{(1+u\bar u)^2}. \ee
In order to reduce it to the system on $\mathbb{H}P^n$, it is
convenient to give the Hamiltonian formulation
 of this system and then perform  the Hamiltonian reduction associated with the Hopf map.
 Precisely, in the Hamiltonian language the free particle system on $\mathbb{C}P^n$
is defined by the triple \be \left(H_0=(g^{-1})_i{}^jp_i{\bar
p}^j,\qquad \omega=dp_i\wedge dz^i+d{\bar p}_i\wedge d{\bar
z}^i,\quad  T^*\mathbb{C}P^{2n+1}\right), \label{ham0}\ee where
$(g^{-1})_i{}^j=(1+z\bar z)(\delta_i^j+{\bar z}_iz^{j})$ are the
components of the inverse Fubini-Study metric (\ref{mcpn}). This
system possesses the $su(2n+2)$ symmetry algebra given by the
generators \eqref{subos1}. These generators define the following
Noether constants of motion \be R_i= p_i +\bar z_i (\bar z\bar
p),\quad J_i{}^j = \imath\left( z^j p_i -\bar z_i \bar p^j\right)+
\imath\delta_i^j\left( (z p) -(\bar z\bar p)\right).
\label{cpniso}\ee We can extend  the transformation \eqref{ctro}
to the canonical one by adding the following transformation rule
for the conjugated momenta: \be p_\mu=\frac{\bar u}{1+u\bar
u}\pi_\mu+\frac{1}{1+u\bar u}\bar\pi_\mu,\quad p_{2n+1}=
p_u-\frac{1}{1+u\bar u}\left(\bar uw^\mu \pi_\mu
-w^\mu\bar\pi_\mu\right).\label{mtr} \ee It is an exercise to
check that the canonical transformation \eqref{ctr},\eqref{mtr}
leads to the following form of the Hamiltonian: \be
H_0=(g^{-1})_\mu{}^\nu\bar P^\mu P_\nu+(1+u\bar u)^2p_u\bar
p_u,\label{ham1} \ee where we introduced the inverse metric to
\eqref{hpncp} \be (g^{-1})_\mu{}^\nu=(1+w\bar
w)(\delta_\mu^\nu+\bar w_\mu w^\nu+w_\mu \bar w^\nu),\label{hpnmi}
\ee and the
 covariant momenta
\be
P_\mu=\pi_\mu-\imath \bar w_\mu\frac{ I_3}{1+w\bar w}-w_\mu\frac{I_+}{1+w\bar w},
\ee
 with the $su(2)$ generators $I_\pm,I_3$ defining the isometries of $S^2$:
\be
I_3=-\imath(up_u-\bar u\bar p_u),\quad I_+=\bar p_u+u^2p_u,\quad I_-=p_u+\bar u^2\bar p_u,\ee
\be
\left\{I_3,I_\pm\right\}=\pm\imath I_\pm,\quad \left\{I_+,I_-\right\}=2\imath I_3,\label{pbu}
\ee
The Poisson brackets between the quantities $P_\mu$ read
\be
\left\{w^\mu,P_\nu\right\}=\delta^\mu_\nu,\quad\left\{P_\mu,P_\nu\right\}=-2\frac{\Omega_{\mu\nu}}{1+w\bar w}I_+,\quad \left\{P_\mu,\bar P^\nu\right\}=\imath\frac{\delta_\mu^\nu I_3}{(1+w\bar w)^2}.
\label{pb1}\ee
Besides, we have
\be
\{P_\mu,I_+\}=\frac{\bar w_\mu I_+}{1+w\bar w},\quad\{P_\mu,I_-\}=-\frac{\bar w_\mu I_-}{1+w\bar w}-2\imath\frac{w_\mu I_3}{1+w\bar w},\quad \{ P_\mu,I_3\}=\frac{\imath w_\mu I_+}{1+w\bar w}.
\label{pb3}\ee
Let us also present, for completeness,  the some other relations as well
\be
\{I_3, p_u\}=-\imath p_u,\quad \{I_+,p_u\}=2u p_u,\quad \{I_-,p_u\}=0,
\quad
\{P_\mu,p_u\}=-\frac{p_u}{1+w\bar w}\left(\bar w_\mu+2u w_\mu\right).
\ee

It is easy to see that the Casimir operator of these  $su(2)$
generators  is precisely the Hamiltonian of a free particle moving
on $S^2$, in \eqref{ham1} as a second summand: \be I^2=I_+
I_-+I_3^2=(1+u\bar u)^2p_u\bar p_u.\label{s2iz} \ee Obviously, it
commutes with the Hamiltonian and defines an integral of motion of
the system.

Our goal is to perform the Hamiltonian reduction by this constant
of motion. For this purpose  we should fix its
$(4(2n+1)-1)$-dimensional level surface, by putting \be I^2=s^2,
\ee and then factorize it by the vector flow $\{I^2,\;\}$ to the
$(8n+2)=(2\cdot 4n +2)$ dimensional phase space.

The role of the $8n$ coordinates of the reduced phase space could
be played by $P_\mu, w^\mu$ given by \eqref{mtr}, \eqref{ctro}. To
find the rest two coordinates, we should simply resolve the
condition \eqref{s2iz} preserving Poisson brackets \eqref{pbu}:
\be I_+=s\frac{2x}{1+x\bar x},\quad I_3=s\frac{1-x\bar x}{1+x\bar
x},\quad s\in\mathbb{R},\quad x\in \mathbb{C}. \ee This yields \be
x=\frac{I_+}{s+I_3},\;:\; \left\{x,\bar
x\right\}=\frac{i}{2s}(1+x\bar x)^2,\quad\{P_\mu, x\}=\frac{\bar
w_\mu x}{1+w\bar w}-\imath\frac{x^2 \bar w_\mu}{1+w\bar w},
\label{xpb}\ee
$$
\{P_\mu,{\bar x}\}=-\frac{\bar w_\mu\bar x+\imath w_\mu}{1+w\bar w},\quad \{w^\mu, x\}=\{w^\mu, {\bar x}\}=0
$$
Thus, we have reduced the phase space $T^*\mathbb{C}P^{2n+1}$ to
the $T^*\mathbb{H}P^{n}\times S^2$. The latter defines the phase
space of the ($su(2)$-)isospin particle on $\mathbb{H}P^n$
interacting with the BPST instanton field. Its Poisson brackets
are defined by the relations (\ref{pb1}),(\ref{xpb}), or,
equivalently, by (\ref{pbu}),(\ref{pb1}),(\ref{pb3}).

The Hamiltonian of the free particle on  $\mathbb{C}P^{2n+1}$
given by \eqref{ham0} (or, equivalently, by \eqref{ham1}),
results, upon reduction, to the one on $\mathbb{H}P^n$, \be
H_{\mathbb{H}P^n}=(g^{-1})_\mu{}^\nu \bar P^\mu P_\nu+s^2.
\label{hpnham}\ee So, we reduced the free particle system on
$T^*\mathbb{C}P^{2n+1}$ to the isospin particle on $\mathbb{H}P^n$
interacting with the BPST instanton field. The subset of the
Noether constants of motion  \eqref{cpniso} commuting with
\eqref{s2iz} is reduced on $\mathbb{H}P^n$ and form the
 $Sp(n+1)$ algebra of the isometries of the reduced system.

 These reduced generators are given by the expressions
\be L_\mu{}^\nu=J_\mu{}^\nu+ J^\nu{}_\mu= \imath(w^\nu \pi_\mu-\bar
w_\mu \bar\pi^\nu)+\imath\left(w_\mu\pi^\nu- \bar w^\nu
\bar\pi_\mu\right) \ee \be L_3=J_{2n+1}{}^{2n+1}=
I_3+\frac{\imath}{2}((\pi w)-(\bar\pi\bar w)), \qquad
L_-=R_{2n+1}=(\bar\pi^\mu w_\mu)+I_-, \quad L_+= {\bar L}_- \ee \be
L^\mu=\imath J_{2n+1}{}^\mu-R^\mu=\bar\pi^\mu-((\bar\pi^\nu
w_\nu)-I_-)\bar w^\mu+((\pi w)-\imath I_3)w^\mu \ee

The $2n(2n+1)/2$ generators $L_{\mu\nu}$ and three generators
$L_\pm, L_3$, form the $sp(n)\times sp(1)$ algebra, and these two
sets of generators, together with the $4n$ generators $L_\mu$,
form the $2n^2+5n+3$-dimensional algebra of the isometries of
$\mathbb{H}P^n$, that is $sp(n+1)$: \be
\{L_{\mu\nu},L_{\rho\sigma}\}=-\imath(\Omega_{\mu\rho}L_{\sigma\nu}+\Omega_{\nu\rho}L_{\sigma\mu}+\Omega_{\nu\sigma}L_{\mu\rho}+\Omega_{\mu\sigma}L_{\nu\rho})
\ee \be \{L_+,L_-\}=2\imath L_3,\quad \{L_3,L_\pm\}=\pm \imath
L_\pm,\qquad \quad \{L_{\mu\nu},L_\pm\}=\{L_{\mu\nu},L_3\}=0 \ee
\be \{L^\mu,L_-\}=0,\quad \{L^\mu,L_+\}=\bar L^\mu,\quad
\{L^\mu,L_3\}=\frac{\imath}{2} L^\mu,
\quad \{L^\mu,L_\rho{}^\sigma\}=-\imath \delta^\mu_\rho L^\sigma-\imath L_\rho\Omega^{\mu\sigma}
\ee
\be
\{L^\mu,L^\nu\}=2\Omega^{\mu\nu}L_-,\quad \{\bar L_\mu,\bar L_\nu\}=-2\Omega_{\mu\nu}L_+,\quad \{L^\mu,\bar L_\nu\}=\imath L_{\nu}{}^\mu+2\imath\delta^\mu_\nu L_3
\ee

The Casimir  of $sp(n+1)$  is precisely  the Hamiltonian on
$\mathbb{H}P^n$ \be
H_{\mathbb{H}P^n}=\frac{L_{\mu}{}^{\nu}L_{\nu}{}^{\mu}+4 L^\mu
\bar L_\mu+8(L_+L_-+L_3^2)}{4}+I^2 \ee Hence, we established a
complete correspondence between the $SU(2)$ (classical) Landau
problem on $\mathbb{H}P^n$ and the free particle system on
$\mathbb{C}P^{2n+2}$

\setcounter{equation}{0}
\section{Oscillator}
In the previous section we considered the free particle on the
n-dimensional quaternionic projective space $\mathbb{H}P^n$ and
demonstrated, that the inclusion of $SU(2)$ instanton  preserves
its whole symmetry algebra $sp(n+1)$. In fact $\mathbb{H}P^n$
seems to be a natural candidate on the role of configuration
spaces of the (super)integrable systems interacting with the BPST
instanton field. At least, on these spaces there should exist the
proper generalizations of the systems on $\mathbb{R}^4$ respecting
the inclusion of the BPST instanton. The simplest system of this
sort, besides the free particle,  is the $4n$-dimensional
isotropic oscillator.
 How to  construct its appropriate analog  on $\mathbb{H}P^n$?

Let us consider a more complicated integrable system on
$\mathbb{H}P^n$, that is the generalization of oscillator, given
by the following expression: \be
H_{osc}=H_{\mathbb{H}P^n}+\omega^2_0 w\bar w, \label{osc}\ee where
the first term is simply the free particle Hamiltonian on
$\mathbb{H}P^n$ given by (\ref{hpnham}).

This potential  has been suggested in \cite{lnp,mardoyan} in
analogy with the  earlier constructed oscillator potential on
$\mathbb{C}P^n$. In our opinion, it is deductive to present here
the speculations which lead to the suggestion of above system.
Namely, in \cite{cpnosc} two of the authors constructed the model
of the oscillator on $\mathbb{C}P^n$ requiring that it should have
the hidden symmetries, resulting, in the flat limit, in the
ordinary oscillator on $\mathbb{C}^n$. Such a model was found to
be unique. On $\mathbb{C}P^1$ it was found to be the well-known
Higgs oscillator, while for $n>1$ it was defined by the potential
\be V_{\mathbb{C}P^n}= \omega_0^2 z\bar z, \ee with $z^i$ being
the inhomogeneous coordinates on $\mathbb{C}P^n$.
 Besides the $u(n)$ N\"other constants of motion  defined by the second expression in (\ref{cpniso}),
 this system was found to have hidden constants of motion (for $n>1$) given by the expression
 \be
 I^i_j={\overline R}^iR_j+\omega^2_0z^i{\bar z}_j,
 \label{cpnI}\ee
 where $R_i$ are the translation generators defined by the first expression in (\ref{cpniso})
 Surprisingly, it  was found that the inclusion of
of the constant magnetic field preserves all symmetries (and,
respectively, superintegrability) of the system! Moreover, it was
found, that even on $\mathbb{C}P^1$ this potential is a
distinguished one. Namely, though the system is not
superintegrable in this case, it is exactly solvable, and
preserves the exact solvability property after inclusion of the
constant magnetic field, while the Higgs oscillator on
$S^2=\mathbb{C}P^1$, being a superintegrable system,
 looses the superintegrability (and even the exact solvability) property upon inclusion of a constant magnetic field.
This allowed the authors to call that system
``$\mathbb{C}P^n$-oscillator" for any $n$. It was further studied
in \cite{cpnother}.

Taking in mind, that the potential of the Higgs oscillator on the
n-dimensional sphere (to be more precise, on the real projective
space) reads, in inhomogeneous coordinates \be
V_{\mathbb{R}P^n}=\frac{\omega^2_0y^2}{2}, \qquad
y^i=\frac{u^i}{u^0}, \ee with $u^i,u^0$ being coordinates of the
ambient $\mathbb{R}^n$ space($(u^i)^2 +(u^0)^2=1$), the authors of
\cite{lnp,mardoyan} claimed, that the  oscillator potential on
$\mathbb{H}P^n$ should be given by the same expression, as in the
case of $\mathbb{C}P^n$, with the replacement of inhomogeneous
complex coordinates with quaternionic ones. And this system has to
respect the inclusion of the BPST instanton field. They have
checked this in the simplest case of $\mathbb{H}P^n=S^4$ and found
that it is indeed the case. However, in contrast with the Higgs
oscillator on $\mathbb{R}P^1=S^1$, and with the
$\mathbb{C}P^1(=S^2)$-oscillator, the spectrum of the
$\mathbb{H}P^1$-oscillator (and of its hyperbolic analog
\cite{mardoyan}) system was found
 to be degenerated, which is a precise indication of the existence of hidden symmetries.
 Unfortunately, no explanation of these symmetries has been done there. Moreover, this claim has never been checked for nontrivial
 (higher-dimensional) cases.

Now, let us show, that the Hamiltonian (\ref{osc}), together with
Poisson brackets (\ref{pbu})-(\ref{pb3}) defines a well-defined
oscillator system on $\mathbb{H}P^n$.

It is clear that the added oscillator potential does not commute
with the coset generators $L^\mu,\bar L_\mu$, while the rest of the
$sp(n)\times sp(1)$ generators $L_{\mu}{}^\nu,L_{\pm,3}$ remain as
symmetries of the system. However, the system possesses a set of
hidden symmetries: \be I_\nu{}^\mu=L^\mu \bar L_\nu-\bar L^\mu
L_\nu+\omega^2_0( w^\mu\bar w_\nu-\bar w^\mu w_\nu), \ee which are
constructed by analogy with the corresponding integrals for the
$\mathbb{C}P^n$ oscillator (\ref{cpnI}). These quantities commute
with $L_{\pm,3}$ and  transform linearly with respect to
$L_{\mu\nu}$: \be
\{I_{\mu\nu},L_{\rho\sigma}\}=\imath(\Omega_{\mu\sigma} I_{\nu\rho}
+\Omega_{\rho\mu}
I_{\sigma\nu}-\Omega_{\nu\sigma}I_{\mu\rho}-\Omega_{\nu\rho}I_{\sigma\mu}),\quad
\{L_{\mu\nu},L_{\pm,3}\}=\{I_{\mu\nu},L_{\pm,3}\}=0, \ee However, in
contrast to the case of the $\mathbb{C}P^n$ oscillator, where the
symmetries of the system form a quadratic algebra, in the case of
the $\mathbb{H}P^n$ oscillator the Poisson brackets between the
hidden symmetry generators cannot be expressed through the
combination of $L$ and $I$ and give us a new set of integrals of
motion, which are, already, cubic in momenta: \be
\{I_{\mu\nu},I_{\rho\sigma}\}=\imath(I_{\mu\rho}L_{\nu\sigma}+I_{\nu\sigma}L_{\mu\rho}-I_{\mu\sigma}L_{\nu\rho}
-I_{\nu\rho}L_{\mu\sigma})+\Omega_{\mu\rho}S_{\nu\sigma}+
\Omega_{\nu\sigma}S_{\mu\rho}-\Omega_{\mu\sigma}S_{\nu\rho}-\Omega_{\nu\rho}S_{\mu\sigma}
\ee where \be\label{S} S_{\mu\nu}=2 L_\mu L_\nu L_- +2 \bar
L_\mu\bar L_\nu L_++2\imath L_3(L_\mu\bar L_\nu+L_\nu\bar L_\mu)-
\ee
$$
-\omega^2_0\left( L_{\mu\nu}+2(w_\mu w_\nu L_++\bar w_\mu \bar w_\nu L_--\imath
L_3(w_\mu \bar w_\nu+w_\nu\bar w_\mu))\right)
$$
defines the new set of cubic constants of motion. Their Poisson
brackets yield additional, last, set of constants of motion \be
T_{\mu\nu}=I_-(\bar \pi_\mu w_\nu-\bar \pi_\nu w_\mu)+I_+(\pi_\mu
\bar w_\nu-\pi_\nu\bar w_\mu)-\imath I_3(\pi_\mu w_\nu-\pi_\nu
w_\mu+\bar \pi_\mu\bar w_\nu-\bar \pi_\nu\bar w_\mu)
.\label{last}\ee Let us notice, that the constants of motion
(\ref{S}),(\ref{last})  have no analogs neither in Higgs, no in
$\mathbb{C}P^n$ oscillator models. It seems, that precisely these
constants of motion are responsible for the degeneracy of the
spectrum of the $\mathbb{H}P^1$ oscillator observed in
\cite{mardoyan}.

Finally, let us notice, that repeating the speculations given for
the $\mathbb{C}P^n$ oscillator, we can define the singular version
of the $\mathbb{H}P^n$ oscillator respecting the inclusion of the
BPST instanton field, \be V=\frac{\alpha^2}{w\bar w}+\omega^2_0
w\bar w. \label{sing}\ee The classical and semiclassical analysis
of its $\mathbb{C}P^1$ analog has been done in \cite{aramyan}.

\section{Conclusion}
In this paper we constructed two basic one-particle integrable
systems on quaternionic projective spaces, that are the
"quaternionic Landau problem" on $\mathbb{H}P^n$, i.e. the
particle moving in the presence of an instanton field, and the
"$\mathbb{H}P^n$-oscillator" (interacting with the instanton
field). Both systems are superintegrable: the first one possesses
the $Sp(n+1)$  symmetry algebra,
 while the symmetry algebra of the second one is highly nonlinear, and it still needs to be  calculated.
Note, that both systems can be easily lifted to the ones on the
complex projective space, where the instanton field is "absorbed" in
the spatial coordinates. The obvious next step is to consider the
respective quantum mechanical systems for $n>1$ (for the $n=1$ case
it was considered in earlier works). With the quantum mechanics at
hands, one can consider, e.g. the "quantum Hall effect on
$\mathbb{H}P^n$", in analogy with "quantum Hall effect on
$\mathbb{C}P^n$", considered by Karabali and Nair (see the fourth
reference in \cite{4Hallothers} and \cite{Karabali}). A more
detailed description of the "singular $\mathbb{H}P^n$-oscillator"
defined by the potential (\ref{sing}), and the construction and
proper generalization of the Coulomb system are also in order.
Supersymmetric extensions  of these systems, which are similar to
those on complex projective spaces,  are of special importance
\cite{BN1}. However, the natural desire to obtain them by the
Hamiltonian reduction from the complex projective space seems to be
technically irrelevant \cite{Kozyrev},
so that one should try to do it in a less obvious way.\\

{\large Acknowledgements.} We thank  Tigran Hakobyan and David
Karakhanyan  for useful  comments. This work was supported  by the
Armenian State Committee of Science  grant 11-1c258, by the ANSEF
grant 2908, by RFBR grants 11-02-01335, 12-02-00517, and
11-02-90445-Ukr as well as by ERC Advanced Grant no. 226455,
â€œSupersymmetry, Quantum Gravity and Gauge Fieldsâ€
(SUPERFIELDS).

\end{document}